\documentclass{ws-ijmpb}

\begin{document}

\markboth{S. K. Mishra {\it et al.}}
{Size-dependent magnetization fluctuations in NiO nanoparticles}

%
\catchline{}{}{}{}{}
%

\title{ Size-dependent magnetization fluctuations in NiO nanoparticles }

\author{Sunil Kumar Mishra \and V. Subrahmanyam }

\address{Department of Physics, Indian Institute of Technology\\
 Kanpur-208016, India,\\
sunilkm@iitk.ac.in }
%
%
\maketitle
\begin{history}
\end{history}
\begin{abstract}
The finite size and surface roughness effects on the magnetization of NiO nanoparticles is
investigated. A large magnetic moment arises for an antiferromagnetic nanoparticle due to
these effects. The magnetic moment without the surface roughness has a
non-monotonic and oscillatory dependence on $R$, the size of the particles, with the amplitude of
the fluctuations varying linearly with $R$. The geometry of the particle also matters a lot in the calculation of the net magnetic moment. An oblate spheroid shape particle shows an increase in net magnetic moment by increasing oblateness of the particle. However, the magnetic moment values thus calculated
are very small compared to the experimental values for various sizes, indicating that the
bulk antiferromagnetic structure may not hold near the surface.  We incorporate the surface
roughness in two different ways; an ordered surface with surface spins inside a surface roughness shell aligned due to an internal field, and a disordered surface with randomly oriented spins inside surface roughness shell.
Taking a variational approach we find that the core interaction strength is modified for nontrivial values of $\Delta$ which is a signature of multi-sublattice ordering for nanoparticles. The surface roughness scale $\Delta $ is also showing size dependent fluctuations, with an envelope decay 
$\Delta\sim R^{-1/5}$.
The net magnetic moment values calculated using spheroidal shape and ordered surface are close to the experimental
values for different sizes.
\end{abstract}

\section{Introduction}
\label{intro}
Antiferromagnetic nanoparticles have been receiving a refreshed research attention over the last few years. 
These are considered as better candidates for exhibiting the magnetization reversal by quantum tunneling \cite {stamp}, due
to their small magnetic moment as compared to the ferromagnetic nanoparticles. 
The magnetic properties of nanoparticles are dominated by finite-size effects, and the surface anomalies such as
surface anisotropy and roughness.\cite{fiorani,kodama,kodama1} As the particle size decreases, the fraction of
the spins lying on the surface of a nanoparticle increases, thus, making the surface play an important role. The
reduced coordination of the surface spins causes a symmetry lowering locally, and leads to a surface anisotropy,
that starts dominating as the particle size decreases. 
Thus, an enhancement of surface and interface effects make the antiferromagnetic nanoparticles an interesting 
area of research.\cite{fiorani,kodama,kodama1,winkler,tiwari,yi,winkler1,morales,behl,tomou,bhowmik,jagodic}

Nickel Oxide (NiO) has been considered as a prototype for antiferromagnetism, as it is one of the first few materials
in which antiferromagnetism was studied.\cite{shull}
One of the first serious concerns with NiO nanoparticle was, evidenced from the experimental study of Richardson
and Milligan,\cite {richard1} that these nanoparticles show a large magnetic moment as the size becomes smaller than
$100 \rm{nm}$, apart from  
anomalous behavior  of the magnetic susceptibility.
It was also found that the exchange coupling between the surface spins and the antiferromagnetic core spins causes 
an exchange bias phenomenon in these finite-sized particles. This phenomenon is responsible for 
the observed shifted hysteresis loop after field cooling in NiO nanoparticles.\cite{kodama} 
This interface effect is very much size dependent. 
A large loop shift $\left( \ > 10 KOe \right) $ and coercivities at low temperature has been reported for the intermediate sized particles ($22 \rm{nm}-31 {nm}$).\cite{kodama,kodama1,makhlouf} 

Winkler {\it et al} \cite{winkler} reported that for $3 \rm{nm}$ particles the magnetization curves are reversible 
above $T \sim 170 K$, but a hysteresis behavior is observed at lower temperatures. According to their observation, a large surface
anisotropy is responsible for the anomalies in the shape of the hysteresis loop at low temperatures. They found that with
a decrease in the temperature, a progressive blocking of the core particle moments starts off, and it is followed by 
a growth of spin clusters at the particle's surface below $40 K $, and finally their collective freezing in a 
cluster glass-like state at $15 K$.

The net magnetic moment of antiferromagnetic nanoparticles has been a subject of research interest from a long time. 
N\'eel in 1961 suggested \cite{neel} that fine particles of antiferromagnetic materials exhibit weak ferromagnetism 
and superparamagnetism. He argued that the permanent magnetic moment in these antiferromagnetic fine particles is due 
to incomplete magnetic compensation between the atoms on the two sublattices `A' and `B', which are identical in every
respect, except that the atomic moments in B sublattice are antiparallel to that in A sublattice. N\'eel considered
three general cases as shown in Ref.~\cite{richard2}. If the uncompensation of spins occurs randomly in a particle, then the number
of uncompensated spins $p$ will vary as $p\sim n^{\frac{1}{2}}$, where $n$ is the number of spins. If the spins are arranged 
in such a way that the ordered structure consists of odd number of ferromagnetic planes of A and B atoms, then 
$p\sim n^{\frac{2}{3}}$. Finally, if each plane consists of equal numbers of A and B atoms and the structure consists
of incomplete top and bottom planes, then we would have $p\sim n^{\frac{1}{3}}$. Richardson {\it et al} \cite{richard2}
showed that  $p\sim n^{\frac{1}{3}}$ from the size dependence of susceptibility in NiO nanoparticles. Thus, according to 
the N\'eel's model, the magnetic moment $\mu$ for NiO nanoparticles varies as $ \mu \sim n^{\frac{1}{3}} \mu_{Ni^{2+}} $.
Weak ferromagnetism was later confirmed by experiments \cite {schuele} on fine particles of NiO. For extremely
fine particles they reported the behavior to be superparamagnetic. However, Tiwari {\it et al} \cite{tiwari} 
argued that the NiO nanoparticles behave like a superspin glass, 
which is attributed to a surface spin disorder. Some authors accredited the large magnetic moment in NiO nanoparticle
to nonstoichiometry, an existence of small superparamagnetic metallic nickel clusters in NiO particle or the presence of
$Ni^{3+} $ ions within the NiO lattice.\cite{jacob} However Richardson {\it et al} \cite{richard2} confirmed that the
presence of $Ni^{3+} $ in NiO do not contribute significantly to the magnetic moment of NiO nanoparticles.
More recently Yi {\it et al} \cite{yi} investigated the size dependent magnetic properties of NiO nanostructures using experimental and first principle study. They reported that NiO clusters with a size upto $1 \rm{nm}$ indicate
ferromagneticlike interactions with high magnetizations, and NiO nanocrystals with a
particle size over $2 \rm{nm}$ possess uncompensated magnetization.  

The magnetic moment per particle for NiO has been investigated experimentally by Kodama {\it et al}.\cite {kodama} 
From extrapolation of $5 K$ magnetization curves from a large magnetic field to zero field, they found 700 $\mu_{B}$ per
particle for particles of size $15 \rm{nm}$, while the N\'eel's two-sublattice model \cite {richard2} predicts a magnetic moment
of about 80 $\mu_{B}$. For the particles of size $3 \rm{nm}$, Winkler {\it et al} \cite{winkler} experimentally found the magnetic
moment per particle to be 500 $\mu_{B}$, whereas for this particle size, the N\'eel's model predicts a magnetic moment of
20 $\mu_{B}$.

This discrepancy, between the magnetic moments experimentally observed and those predicted by the 
N\'eel's two-sublattice model has been a serious question from a long time. 
 Kodama {\it et al} \cite {kodama} have shown from numerical modeling that a reduced symmetry on the surface of the nanoparticle
actually causes a fundamental change in the magnetic order which results in a multi-sublattice structure.
Monte Carlo studies for antiferromagnetic nanoparticles  by Zianni {\it et al} \cite {trohidou} also reveals a distinct magnetic role of surface and core spins.
 Recently it has been pointed out that the roughness at the  
surface layer gives rise to higher magnetic response for the surface spins than the core spins.\cite{labarta,labarta1,labarta2,labarta3}

In view of these studies, we investigate the large magnetic moment in NiO nanoparticle by invoking a different ordering for surface spins than the bulk N\'eel state ordering for core spins.

\begin{figure}[bt]
\centerline{\psfig{file=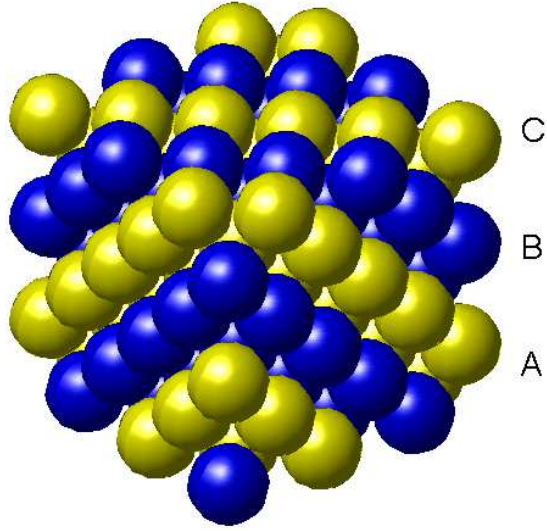,width=3.65in}}
\vspace*{8pt}
\caption{The magnetic configuration of bulk NiO where closed-packed ferromagnetic sheets of spins are stacked 
antiferromagnetically along the direction perpendicular to the sheet which is $\left\langle  111\right\rangle$ direction for bulk NiO. The yellow (white) spheres denote up spins and the blue (black) spheres denote down spins.}
\label {fig2}
\end{figure}

The outline of the present manuscript is as follows: In Sect.~\ref{model} we discuss a model for bulk NiO, followed by a discussion on the finite-size
effect in the magnetization of NiO nanoparticles using spherical as well as spheroidal geometries in Sect.~\ref{finite}. We will discuss surface effects and ordering of surface spins beyond N\'eel state ordering for nanoparticles in Sect.~\ref{surface} and Sect.~\ref{conclusion} is devoted to  conclusions.

\section{The model}
\label{model}
The crystal structure of bulk NiO has been comprehensively investigated in the literature using x-ray diffraction method
\cite{rooksby,slack}. It has been found to be face centered cubic (fcc), with a N\'eel temperature
of
$523 K$. Each Ni atom has twelve nearest neighbors and six next-nearest neighbors. 
 The lattice parameter has been found  \cite{bartel} to be $4.1758 \AA{}$ at $297 K$ and $4.1705 \AA{}$ at 
$T \rightarrow 0$ K

The magnetic structure of NiO has been well established to be fcc-II by the work of  Shull {\it et al} \cite {shull} and further by Roth {\it et al}.\cite{roth1,roth2,roth3} 
The atomic spins are stacked ferromagnetically in $\left( 1 1 1 \right)$  plane but aligned
antiferromagnetically in $\left\langle  1 1 1 \right\rangle$ directions. The direction of alignment of the spin moments has been found to be $\left\langle  1 1 \bar{2}\right\rangle$ directions. The magnetic
configuration of bulk NiO is shown in Fig. \ref{fig2}.
\begin{figure}[bt]
\centerline{\psfig{file=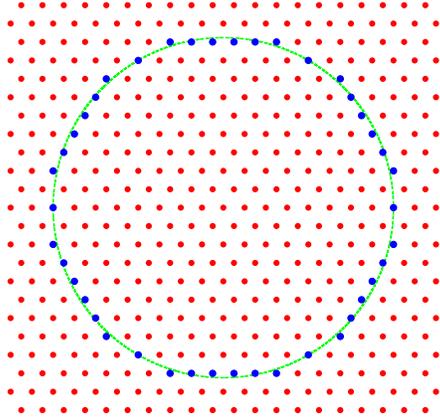,width=3.65in}}
\vspace*{8pt}
\caption{A portion of $\left( 1 1 1\right)$ plane of the bulk NiO is shown. A circle, whose centre is one of the lattice points itself is drawn. The circle shown is a cross section of NiO nanoparticle. The lattice points on the perimeter of the circle are highlighted. Some of the lattice 
points are lying just inside the perimeter and some just outside. These lattice points are 
responsible for the fluctuations.}
\label {fig3}
\end{figure}

 The neutron diffraction
studies by Hutchings \textit{et al} \cite{hutchings} confirmed that the predominant interaction in NiO is a large
next-nearest neighbor antiferromagnetic exchange interaction $J_{nnn}=221 K$ linked by $180^{0}$ superexchange path 
$ Ni^{2+}-O^{2-}-Ni^{2+} $. The nearest-neighbor interaction is linked by $90^{0}$ path  $ Ni^{2+}-O^{2-}-Ni^{2+} $, which
is much smaller in strength. Due to the lattice contraction, there is a slight difference in the exchange interaction 
between the nearest neighbors in the plane, $J_{nn}^- = 16.1 K$, and between nearest neighbors out of the plane, $J_{nn}^+ = 15.7 K$.
 Hutchings \textit{et al} \cite{hutchings} also used an orthorhombic
form for the anisotropy $E_i^A=K_1(\vec{s}_i.\hat{x})^2+K_2(\vec{s}_i.\hat{z})^2$ with $\hat{x}$ is the easy axis direction $\langle112\rangle$ and $\hat{z}$ is the hard axis direction $\langle111\rangle$, where $\vec{s}_i$ is the atomic spin at site $i$. The anisotropy constants are gives as $K_1=1.13 K$ and $K_2=.06 K$. Since $K_2$ is much smaller than $K_1$, we use $K_1$ as anisotroy constant and $\hat{z}$ as anisotropy axis.

The spins in the
NiO interact via Heisenberg exchange interaction.
The Hamiltonian of the system in the presence of an external magnetic filed $H$ is
\begin{eqnarray}
 \mathcal{H}&=& J_{nnn}\sum_{\langle ij \rangle} \vec{s}_i.\vec{s}_j -J_{nn}^-\sum_{\langle ij \rangle} \vec{s}_i.\vec{s}_j +J_{nn}^+\sum_{\langle ij \rangle} \vec{s}_i.\vec{s}_j \nonumber \\
&-& K_1\sum_i (\vec{s}_i.\hat{z})^2-\vec{H}.\sum_i\vec{s}_i.
\label{energy}
\end{eqnarray}
The first term represents the dominant antiferromagnetic next-nearest neighbor exchange energy. These next-nearest neighbors lie in the adjacent planes just above and just below the plane consisting the spin $s_i$, {\it e.g.} for each spin in the plane B in Fig. \ref{fig2}, three of the six next-nearest neighbors lie in plane A while other three lie in plane C. The second term represents the ferromagnetic nearest neighbor exchange energy which determines the  interaction of $s_i$ with six of the twelve nearest neighboring spins lying in the same plane as spin $s_i$. The third term  is antiferromagnetic nearest neighbor interaction energy which represents the antiferromagnetic interaction of $s_i$ with six nearest neighbors lying in the planes other than the plane containing spin $s_i$,  {\it e.g.} each spin in the plane B, has three nearest neighbors in plane A and three in plane C. The fourth term represents the uniaxial anisotropy energy and the last term is the Zeeman energy.\\
The most dominant term in the above Hamiltonian is the first term which supports antiferromagnetic order. Thus, in the bulk we have a N\'eel state ordering where spins are stacked ferromagnetically in $\left( 1 1 1 \right)$  plane but aligned antiferromagnetically in $\left\langle  1 1 1 \right\rangle$ directions. Though the N\'eel state in the bulk has zero magnetization, however, as the size of the particle becomes smaller, the N\'eel state ordering shows a magnetization, since the spins lying near the surface do not cancel out. We will examine below the finite size effect on the N\'eel state magnetization.

\section{Finite-size effects in the N\'eel state} 
\label{finite}
We consider various geometries for antiferromagnetic NiO nanoparticles.
The crystal structure of NiO nanoparticles is the same as that of bulk NiO, except that the unit cell is slightly enlarged.\cite{behl1,mita} 
The spherical geometry of NiO nanoparticle consists of circles stacked with decreasing radius on both sides of the equatorial great circle. These circles are circular cross-sections of $\left( 111\right)$ planes of NiO. The lattice sites in these circular planes are arranged in a triangular lattice structure. We show a part of a  $\left( 111\right)$ plane in Fig. \ref{fig3}. The separation between two neighboring planes is 
$\delta= 2 a/ \sqrt{3}$,
where $a$ is the triangular lattice parameter which is related to the cubic lattice parameter $a_{0}$ as  $ a= a_{0}/\sqrt{2}$.
These circular planes in a NiO nanoparticle are stacked in a sequence
 A-B-C-A-B-C $\cdots$ as shown in Fig. \ref{fig2}, where  A, B, C  planes are
 distinguished from each other by a shift of their centers from the origin. We
 label $0, 1, 2$ to the lattice points in the successive planes A, B, C. The position vectors of the center of planes $A, B, C$ can be given (using cartesian unit vectors $\hat{i}$ and $\hat{j}$) as
$\vec{r}_{0}=0$, $\vec{r}_{1}=a(\frac{\hat{i}}{2} + \frac{1}{2\sqrt{3}}\hat{j})$, and 
$\vec{r}_{2}=a(\frac{\hat{i}}{2} - \frac{1}{2\sqrt{3}}\hat{j})$ respectively.
We can write the three-dimensional position vector of the lattice sites in $l^{th}$ plane labelled by integers $m$, $n$ and $l$ 
\begin{figure}[bt]
\centerline{\psfig{file=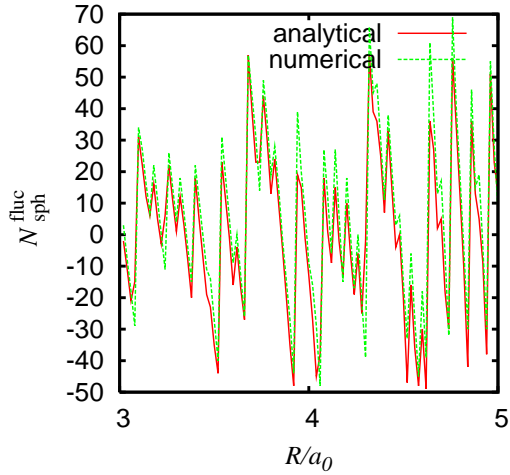,width=3.65in}}
\vspace*{8pt}
\caption{The fluctuation in the total number of spins in a sphere of radius $R$ is
shown. The solid line shows the fluctuation $N_{\rm{sphere}}^{fluc}$ obtained from analytical expression Eq.
(\ref{nsph}) for the terms  up to $G=400$ and Eq. (\ref{spflc}), while the dotted line is the exact
numerical counting of the fluctuation from Eq. (\ref{nsphr}) and Eq. (\ref{spflc}).}
\label {numsp}
\end{figure}
(using the 
cartesian unit vectors $ \hat{i}, \hat{j}, \hat{k} $) as
\begin{eqnarray}
 \vec{r}_{mnl}=(m+\frac{n}{2})a\hat{i}+\frac{\sqrt{3}n}{2}a\hat{j}+l\delta \hat{k}.
\end{eqnarray}
If $N_0$, $N_1$ and $N_2$ are the total number of lattice sites, counting from the planes of type A, B and C  respectively, then the total number of lattice sites within a sphere of radius $R$   will be  given as,
\begin{eqnarray}
 N_{\rm{sphere}}(R)=\displaystyle\sum_{I=0}^{2} N_I(R),
\label{nsphr}
\end{eqnarray}
where,
\begin{eqnarray}
N_I(R) = \displaystyle\sum_{\underset{ l\equiv I\left(Mod 3\right) }{\vec{r}_{mnl}}} \Theta\left( R-\vert \vec{r}_{mnl}-\vec{r}_I\vert\right),
\label{eqn19}
\end{eqnarray}
 $\Theta$ represents Heaviside step function \cite{abramowitz} and $\vec{r}_I$, $(I=0, 1, 2)$, has been discussed above.
We transform the above equation using the Poission sum formula \cite{lighthill} as
\begin{eqnarray}
N_{I}(R) &=& \frac{1}{a^3} \sum_{p,q,w} \int e^{2\pi i (x' p + y' q + z' w)}\nonumber\\
 &\times& \Theta\left( R-\vert\vec{r'}-\vec{r}_I \vert\right) d^{3}r'.
\label{numbcalc}
\end{eqnarray}
Thus, the total number of spins can be written as
\begin{eqnarray}
N_{\rm{sphere}}(R)=\frac{16 \pi}{3} \left( \frac{R}{a_0}\right)^3\displaystyle\sum_{I=0}^{2} \sum_{\lbrace\vec{G}\rbrace} \cos\left( \vec{G}.\vec{R}_I\right)  \frac{j_1\left( GR \right)}{G R},
\label{nsph}
\end{eqnarray}
 where $\vec{G}=\frac{4\pi}{\sqrt{3} a}\left[p \frac{\sqrt{3}}{2}\hat{i}+(q-\frac{p}{2}) \hat{j}+\frac{w}{2\sqrt{2}}\hat{k}\right]$ is a three dimensional reciprocal lattice vector, labeled by three integers $p$, $q$ and $w$, and $\vec{R}_I=\vec{r}_I+I\delta \hat{k}$.
 $j_1$ is a spherical Bessel function of order one \cite{abramowitz}. 

Due to the oscillatory behavior of the Bessel function, $N_{\rm{sphere}}(R)$ varies nonmonotonically with the particle
size $R$, and the wavelength of oscillations goes as $1/G$. 
Thus, the longest wavelength mode $\vec{G}=0$ in the above gives the smooth contribution as
$N_{\rm{sphere}}^{Bulk}=\frac{16}{3}\pi\left( \frac{R}{a_0}\right)^3$,
while the terms with $\vec{G} \neq 0$ represent oscillatory fluctuations.
The fluctuation $N_{\rm{sphere}}^{fluc}$ in the 
total number of spins in the sphere of radius $R$ can be obtained from 

\begin{figure}[bt]
\centerline{\psfig{file=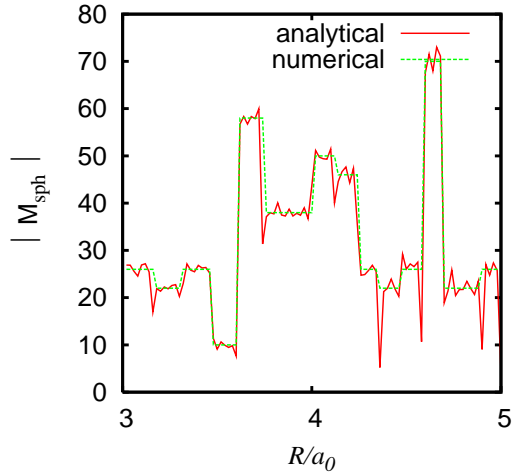,width=3.65in}}
\vspace*{8pt}
\caption{The total magnetic moment $M_{\rm{sphere}}(R)$ of the N\'eel-state for the sphere of radius $R$. The solid line is $M_{\rm{sphere}}(R)$ obtained from analytical expression Eq. (\ref{magf}) for the terms up to $\vert\vec{G}\vert = 400 $ whereas the dotted line shows the numerical counting using Eq. (\ref{mnum1}) and Eq. (\ref{mnum2}).}
\label {fig8}
\end{figure}

\begin{eqnarray}
  N_{\rm{sphere}}=N_{\rm{sphere}}^{\rm{Bulk}}+N_{\rm{sphere}}^{\rm{fluc}}.
\label{spflc}
\end{eqnarray}
From Eq. (\ref{nsph}), and the asymptotic behavior of the spherical Bessel function, $j_1(x)\sim 1/x$, we can see that
the amplitude of oscillatory fluctuations varies as, 
\begin{equation}
N_{\rm{sphere}}^{\rm{fluc}}\sim R.
\end{equation}
Hence the next to leading order term in the number of spins within a sphere
goes as $R$ rather than $R^2$, which we could
have expected from a random-walk argument, viz. the amplitude of the fluctuations is proportional
to the square root of the the number of points on the boundary, here the spherical surface.
 The amplitude of fluctuation varies as $1/G^2$. Thus larger $G$ values give smaller 
contribution to the amplitude of fluctuation. Hence, we can get a good approximation to the 
total 
number of spins by retaining only few terms in the sum in Eq. (\ref{nsph}).
In Fig. \ref{numsp}, we have plotted the fluctuation in the number of spins as obtained from Eq. (\ref{spflc}) for the terms up to $\vert\vec{G}\vert=400$
along with that obtained from the exact numerical counting of spins.
 
\subsection{Magnetization fluctuations of the N\'eel state}
The bulk two-sublattice N\'eel magnetic structure of NiO requires us to assign all the spins in a circular
plane of the NiO nanoparticle to be either +1 or -1. Thus we assign the circular planes to be +1 and -1
alternately, corresponding to ferromagnetic sheets of spins with alternating polarization stacked along $\left\langle 1 1 1\right\rangle$  direction
in the FCC lattice. Since, we have three different types of circular planes, and the circular planes are stacked as
A-B-C-A-B-C  
$\cdots$, as shown in Fig. \ref{fig2}, all the spins in each of A (or B or C) 
type of planes will have either +1 or -1 value depending on the location of the plane along the stacking direction.
The total magnetic moment 
$M_{\rm{sphere}}(R)$ of  NiO spherical particles of size $R$ can be found by summing the magnetic moment of 
all the circular planes.
Following Hutchings {\it et al} \cite{hutchings}, we assume that 
each $Ni^{2+}$ spin has a magnetic moment of $2{\mu}_{\rm{B}}$. 
Thus, we can write the total magnetic moment for the spherical particle as
\begin{eqnarray}
M_{\rm{sphere}}(R) = 2 {\mu}_{\rm{B}} \displaystyle\sum_{I=0}^{2} M_I(R),
\label{mnum1}
\end{eqnarray}
where,
\begin{eqnarray}
M_I(R)= \displaystyle\sum_{\underset{l\equiv I\left(Mod 3\right) }{\vec{r}_{mnl}}} \left(-1 \right) 
^{l}\Theta\left( R-\vert \vec{r}_{mnl}-\vec{r}_I\vert\right).
\label{mnum2}
\end{eqnarray}
 Applying the Poisson sum formula and proceeding analogously as we did above,
\begin{eqnarray}
 M_I(R) &=& \frac{4}{3\sqrt{3}\delta {a_0}^3} \displaystyle\sum_{\lbrace\vec{g},w\rbrace} 
\cos\left(
\vec{g}.{\vec{r}}_I\right) \cos \left( \frac{2 I \pi w}{3}\right)\nonumber\\
&\times& \int e^{i\vec{G}.\vec{r'}} e^{\frac{i\pi z}{\delta}} \Theta\left( R-r'\right)
d^{3}r',
\label{eqn31}
\end{eqnarray}
we evaluate $M_I(R)$ and thus the total magnetic moment in the units of Bohr magneton $(\mu_{\rm{B}})$  as
\begin{eqnarray}
M_{\rm{sphere}}(R) &=& \frac{32\pi}{3} \frac{R ^{2}}{{a_0}^3}\displaystyle\sum_{I=0}^{2}
\displaystyle\sum_{\lbrace\vec{g},w\rbrace}\cos\left( \vec{g}.{\vec{r}}_I\right) \nonumber\\
&\times&\cos \left( \frac{2 I \pi w}{3}\right)
\biggl( G^2 + 3{\pi}^2 
+2\sqrt{3}\pi G_z\biggr )^{-\frac{1}{2}} \nonumber\\
&\times& j_1\left\lbrace R\left( G^2+3{\pi}^2 
+2\sqrt{3}\pi 
G_z\right )^{\frac{1}{2}} \right\rbrace,
\label{magf}
\end{eqnarray}
where
$\vec{G}=\vec{g}+G_{z} \hat{k}$ and $ G_z = \frac{2 \pi}{3\delta} w$.
The total magnetic moment $M_{\rm{sphere}}$ displays oscillations as a function of the particle size, and the wavelength of
oscillations goes as $1/\left( G^2+3{\pi}^2+2\sqrt{3}\pi G_z\right )^{\frac{1}{2}} $. Unlike $N_{\rm{sphere}}$, which
had a smooth part ($\vec G=0$) and oscillatory terms ($\vec G\ne 0$) (see Eq. (\ref{nsph})), all the terms in the above 
Eq. (\ref{magf}) for the total magnetic moment display oscillations. In fact, all the terms have a similar asymptotic behavior.
Using the asymptotic behavior of the Bessel function, the amplitude of the fluctuations in $M_{\rm{sphere}}$ can be shown to
vary as, 
\begin{equation}
M_{\rm{sphere}}\sim R.
\end{equation}
The contribution from the longest wavelength mode $\vec{G}=0$ to the magnetic moment can be written as
\begin{eqnarray}
M_{\vec G=0}= \frac{32}{\sqrt{3}}\left( \frac{R}{a_0}\right)^2 j_1\left( \sqrt{3} \pi \frac{R}{a_0}\right).
\end{eqnarray}
The terms with $(\vec{G} \ne 0)$ represent the fluctuations in the total magnetic moment on various length scales.
The magnetic moment,
as obtained from Eq. (\ref{magf}) for the terms up to $\vert\vec{G}\vert = 
400 $ and as obtained from exact numerical counting is plotted with particle size in Fig. \ref{fig8}.
We find that the net magnetic moment is not as large as seen from experiments. For example, for the
particles of diameter $3 \rm{nm} $, we find the magnetic moment to be $26 
{\mu}_{\rm{B}}$, which is too small compared to experimental value $500 {\mu}_{\rm{B}}$ \cite{winkler}.
Also, for the particles of diameter $15 \rm{nm}$, we find 
the magnetic moment to be $112 {\mu}_{\rm{B}}$, whereas experimental investigation reports $700 {\mu}_{\rm{B}}$ \cite{kodama}. Thus, a net magnetic moment due to N\'eel-state
 does not quantify the large magnetic moment experimentally observed in NiO nanoparticles.

In a realistic situation the shape of NiO nanoparticles is not a perfectly spherical. An oblate spheroid (with high oblateness) or a platelet-shaped geometries has been reported by experiments.\cite{kodama,behl1}
Thus we again calculate magnetization for an spheroidal geometry of nanoparticles. The result of calculations for an oblate spheroid  with polar radius $R/\epsilon$ and equatorial radius $R$, shown in the Appendix, is
\begin{eqnarray}
M_{\rm{spheroid}}(R) &=& \frac{32\pi}{3\epsilon} \frac{ R ^{2}}{{a_0}^3}\displaystyle\sum_{I=0}^{2}
\displaystyle\sum_{\lbrace\vec{g},w\rbrace}\cos\left( \vec{g}.{\vec{r}}_I\right) \nonumber\\
&\times&\cos \left( \frac{2 I \pi w}{3}\right)
\biggl( G^2 + \frac{3{\pi}^2}{{\epsilon}^2}
+2\frac{\sqrt{3}\pi G_z}{\epsilon}\biggr )^{-\frac{1}{2}} \nonumber\\
&\times& j_1\left\lbrace R\left( G^2+\frac{3{\pi}^2}{{\epsilon}^2} 
+\frac{2\sqrt{3}\pi 
G_z}{\epsilon}\right )^{\frac{1}{2}} \right\rbrace,
\end{eqnarray}
and the magnetic moment for the longest wavelength mode $\vec{G}=0$ can be written as
\begin{eqnarray}
M_{\vec G=0}= \frac{32}{\sqrt{3}}\left( \frac{R}{a_0}\right)^2 j_1\left( \frac{\sqrt{3} \pi}{\epsilon} \frac{R}{a_0}\right).
\end{eqnarray}
Using this geometry, we find that the net magnetic moment value increases a little bit but again it is not comparable to experimental values.  for $3 \rm{nm} $ particles, the magnetic moment improves to  
$34{\mu}_{\rm{B}}$ for $\epsilon=2$; $40{\mu}_{\rm{B}}$ for $\epsilon=3$; and $70{\mu}_{\rm{B}}$ for $\epsilon=4$. These values are still very small as compared to experimental value $500 {\mu}_{\rm{B}}$ \cite{winkler}.
Similarly, for the particles of diameter $15 \rm{nm}$, we find 
the magnetic moment to be  $153 {\mu}_{\rm{B}}$ for $\epsilon=2$, $258 {\mu}_{\rm{B}}$ for $\epsilon=3$, and $308 {\mu}_{\rm{B}}$ for $\epsilon=4$.
Thus we see that increasing the oblateness increases the net magnetic moment of the nanoparticle. But the experimental value is still beyond our reach within the present model.
In order to improve the model, we need to invoke a different ordering for surface spins than the bulk N\'eel-state ordering.
We analyze the effects of roughness on the surface of nanoparticle and introduce a surface anisotropy and a ferromagnetic exchange interaction term for the surface spins in the Hamiltonian. The modified model is discussed in the following section and a variational approach is used to find the optimal thickness of surface roughness.

\section{Surface effects and variational approach}
\label{surface} 
The surface effects dominate the magnetic properties of nanoparticles.\cite{fiorani,kodama1,winkler,zysler} The breakdown of the dominant next-nearest neighbor antiferromagnetic interaction on the
surface of the nanoparticle leads to uncompensated 
spins. These uncompensated spins play a vital role in determining the magnetic behavior of NiO 
nanoparticles. The magnetization reversal study for antiferromagnetic nanoparticles by Zianni {\it et al} using 
Monte Carlo simulation \cite {trohidou} also reveals a distinct magnetic role of surface and core spins.  The broken bonds and defects at the 
surface layer gives rise to high magnetic response of the disordered surface spins than the core spins.\cite{labarta,labarta1,labarta2,labarta3}
This surface effect can be incorporated in two different ways; viz. an ordered surface with surface spins ordering owing to internal field due to core spins, and a disordered surface with spins oriented randomly at the surface which cancel out thus giving zero contribution to the total magnetic moment.
 Due to the surface roughness, the uncompensated surface 
spins can be more easily polarized by a small magnetic field. 

For an ordered surface, we take that the spins inside a surface roughness
shell of thickness $\Delta$ are aligned by a field due to a net core magnetic moment though small. This would enhance the net magnetic moment of nanoparticles.
In this scenario, the core spins within a sphere of size $R-\Delta$
have the bulk antiferromagnetic structure, carrying a magnetic moment of order $R$, as we
calculated in Sect.~\ref{finite}, and the spins within a shell of size $\Delta$ are all polarized, carrying
a magnetic moment of order $R^2$.

For a nanoparticle of radius $R$, we can write the total magnetic moment
$M_{\rm{ordered}}\left(R,\Delta\right)$ as
\begin{eqnarray}
 M_{\rm{ordered}}\left(R,\Delta \right)&=& \vert M_{\rm{sphere}}(R-\Delta)\vert+2{\mu}_{\rm{B}} \bigl( N_{\rm{sphere}}\left(
R\right)\nonumber\\
&-&N_{\rm{sphere}}\left( R -\Delta\right) \bigr).
\label{mtot}
\end{eqnarray}
In the above, the first term is due to the core N\'eel-state magnetic moment, and the second term represents the surface roughness effect.
Here, the core spins within a 
sphere of radius $R-\Delta$ have the bulk magnetic structure. 
The spins within the shell of thickness $\Delta$ are aligned, each spin 
contributing a magnetic moment of $2{\mu}_{\rm{B}}$. 
Since $M_{\rm{sphere}}$ in the Eq. (\ref{mtot}) goes as $R$ 
while surface roughness terms as a whole is proportional to $R^2$, the total magnetic moment has a leading term going as $R^2$, if the shell thickness
is independent of the size. 

\begin{figure}[bt]
\centerline{\psfig{file=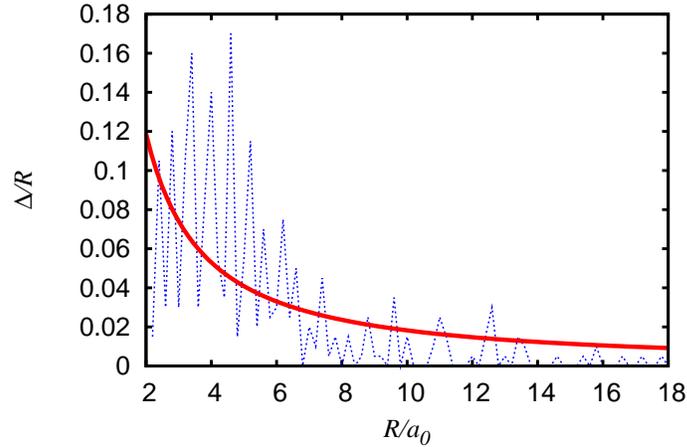,width=3.65in}}
\vspace*{8pt}
\caption{The surface roughness thickness $\Delta$ is plotted with the particle size. Dotted line shows the optimized $\Delta$ obtained from Eq. (\ref{var1}). The fluctuations in $\Delta/R$ show an envelope decay, $\Delta/R\sim .26R^{-6/5}$, shown as a solid line.}
\label {delrs}
\end{figure}

We take a variational approach to establish the behavior of $\Delta$ with the particle size.  In the variational approach, we modify the Hamiltonian (Eq. (\ref{energy})) for nanoparticle, by including a ferromagnetic exchange interaction term $-J_s \sum_{\langle ij \rangle} \vec{s}_i.\vec{s}_j $ with coefficient $J_s$ for spins lying in the roughness shell $\Delta$ and a ferromagnetic exchange interaction term like
$ -J_{cs}\sum_{\substack{ i \in {\rm {core}} \\
         j \in {\rm {surface}}}}\vec{s}_i.\vec{s}_j$ with 
coefficient $J_{cs}$ between core and surface spins lying at the interface of 
core and surface. 
Moreover we introduce surface anisotropy term $-K_s\sum_i (\vec{s}_i.\hat{z})^2$ with coefficient $K_s$ for the surface spins which
 prevails over the uniaxial core anisotropy with coefficient $K_1$. Hence we can ignore the core anisotropy term. 
Taking all the interactions into account exactly, we can write energy of the system  as
\begin{eqnarray}
E(\Delta) &=&(\beta-\alpha)N_{\rm{sphere}}(R-\Delta)-\beta N_{\rm{sphere}}(R) \nonumber\\
&-& \sum_{\substack{ i \in {\rm{core}} \\
         j \in \rm{surface}}}\langle \vec{s}_i.\vec{s}_j\rangle,
\label{var1}
\end{eqnarray}
where $\alpha=z_c+\frac{3J_{nn}^+}{2J_{nnn}}+\frac{3J_{nn}^-}{2J_{nnn}} $ and $\beta=z_s\frac{J_s}{J_{nnn}}+\frac{K_s}{J_{nnn}}$.
$z_c$ and $z_s$ are the coordination numbers of core spins and surface spins. We use the value of $J_{cs}$ same as that of $J_{nnn}$.
 In the above equation, we find that the first term and the last term explicitly contain the variational parameter $\Delta$.
 The energy is minimized with respect to  $\Delta$ to get the optimal thickness of the surface roughness shell. We choose the parameters 
to be $\alpha=6.24a_0/R$ and $\beta=6a_0/R$, such that we have a nontrivial $\Delta$ and the total magnetic moment per particle for an assembly of particles with lognormal distribution is close to the experimental value for the corresponding sizes. This indicates a modification in the core interaction strength in nanoparticles, which signifies a deviation from a two-sublattice ordering.
For large enough sizes, $\Delta$ becomes zero and only  contribution to the net magnetic moment is the first term in Eq. (\ref{mtot}).

\begin{figure}[bt]
\centerline{\psfig{file=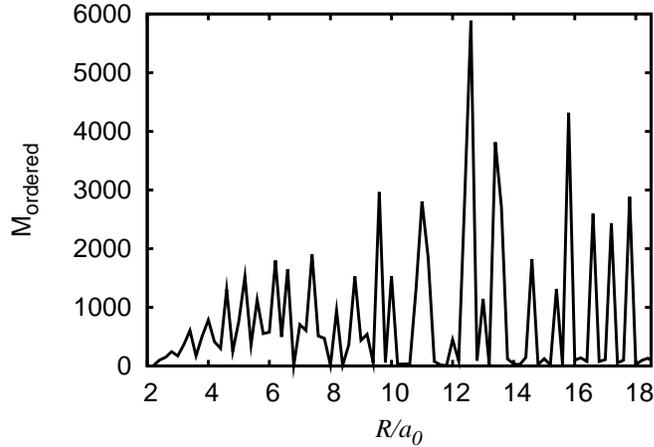,width=3.65in}}
\vspace*{8pt}
\caption{The net magnetic moment as a function of particle size, including the surface roughness and anisotropy, calculated from Eq. (\ref{mtot}). For each size an optimized surface roughness thickness $\Delta$ from Eq. (\ref{var1}) is used.}
\label {fig6a}
\end{figure}

In Fig. \ref{delrs}, the optimal $\Delta/R$ which minimizes energy $E$ in Eq. (\ref{var1}) is plotted with particle size $R$. We find that  $\Delta/R$ shows an oscillatory behavior and the amplitude of oscillations is decreasing with increasing particle size. The best fit of the curve shows that $\Delta \sim .26R^{-1/5}$. In Fig. \ref{fig6a}, we have plotted net magnetic moment for corresponding particle sizes using Eq. (\ref{mtot}). The total magnetic moment displays size-dependent fluctuations, whose amplitude shows a peak at $R\simeq12a_0$. Further increasing the size of the particle lowers the magnetization which occurs due to the lowering of the surface roughness effect i.e., less availability of ordered spins near surface.  A similar behavior has been observed in experiments by Yi {\it et al} \cite{yi} where the magnetization of NiO powder increases with the annealing temperature (grain size) and shows a peak at an annealing temperature $170   
 ^\circ$C. Annealing at higher temperature than $170   
 ^\circ$C leads to a lower magnetization.

\begin{figure*}[t]
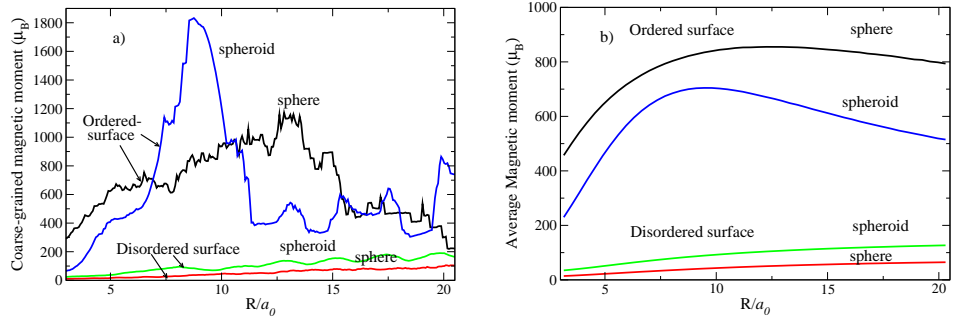

\vspace{.5cm}
$
\begin{array}{cc}
\resizebox{.47\columnwidth}{!}{%
  \includegraphics{figure7a.eps}
}&
\hspace*{0.5cm}
\resizebox{.47\columnwidth}{!}{%
  \includegraphics{figure7b.eps}
}
\vspace*{1.0cm}
\end{array}$
\caption{(a) Coarse-grained magnetic moment versus particle size where coarse-graining is done over a window of size $2.8 a_0$. (b) Averaged magnetic moment versus size where averaging is done over a lognormal size distribution, shown in Eq. (\ref{lognormal}), with a width of the distribution $\sigma=0.5$. In both figures (a) and (b), two different cases ordered and disordered surfaces has been shown. For both cases two different geometries of nanopartcles, a sphere($\epsilon=1$) and a spheroid with $\epsilon=4$ has been shown. An increase in oblateness $\epsilon$ results an enhanced magnetic moment in disordered case, but for ordered surface, magnetic moment for spheroidal particle is less than that of spherical particle. } 
\label{magdist}
\end{figure*}

In the case of a disordered surface, we take a disordered spin structure inside the shell of thickness $\Delta$. In this situation, the net contribution from the disordered surface will be zero and magnetic moment will solely be due to the core spins inside the sphere of radius $R-\Delta$. 
\begin{eqnarray}
 M_{\rm{disordered}}=M_{\rm{sphere}}(R-\Delta).
\label{eqdisord}
\end{eqnarray}
Since $M_{\rm{sphere}}$ shows nonmonotonic oscillatory dependence on particle size $R$, $M_{\rm{disordered}}$ should also show the same behavior with particle size.
In Fig. \ref{magdist}(a) we plot coarse-grained $ M_{\rm{disordered}}$ with particle size $R$ using spherical as well as spheroidal particles. Averaging is done over a  window size of $2.8 a_0$. In the same curve we have also shown ordered surface case. The net magnetic moment increases with increasing oblateness of the spheroidal nanoparticles. In the same Fig. \ref{magdist}(a) we have shown ordered surface case for both geometries, spherical($\epsilon=1$) and spheroidal with $\epsilon=4$. In both the geometries, the coarse-grained magnetic moment shows nonmonotonic oscillating behavior as a function of the size. Though the peak value in spheroid case is more than the sphere, we see a sharp rise and fall of magnetic moment just before and just after the peak ($\approx 9 a_0$) in spheroid case which is different than sphere where the magnetic moment slowly increases with particle size, a plateau is seen for intermediate range, and for bigger sizes it starts decreasing with size. 

The systems of magnetic nanoparticles in experimental studies are in general polydisperse. The shape and size of the particles are not well known but the particle size distribution is often found to be lognormal \cite{buhrman}. 
We consider the system consisting of lognormally distributed, widely dispersed nanoparticles, hence non interacting among each other. 
 The weight of a given size (radius $R$) of nanoparticles is given by a lognormal distribution, 
\begin{eqnarray}
P(R/a_0) = \frac{1}{\sigma( R/a_0) \sqrt{2\pi}} e^{-\frac{(ln(R/a_0)-\mu)^2}{(2\sigma^2)}}.
\label {lognormal}
\end{eqnarray}
The characteristic parameters of the distribution are chosen to be $\mu=ln(\bar{R}/a_0)$ and $\sigma=0.5$. Using this distribution, the magnetic moment for ordered-surface spherical nanoparticle of diameter $3\rm{nm}$ works out to be $510 \mu_{\rm{B}}$ which is quite close to the experimental value $500 {\mu}_{\rm{B}}$ \cite{winkler}. For the ordered-surface oblate spheroidal nanoparticle (oblateness $\epsilon=4$) of same size this value is $297 {\mu}_{\rm{B}}$, which is a little less than the experimental value. Using the same size distribution for disordered-surface spherical nanoparticle and spheroidal nanoparticle, we find net magnetic moment $18{\mu}_{\rm{B}}$
and $40{\mu}_{\rm{B}}$ respectively, which is very small as compared to experimental value. Similarly, for a distribution of particles with mean size $15\rm{nm}$, our calculation of net magnetic moment for ordered-surface spherical particle is $820 {\mu}_{\rm{B}}$, and for order-surface spheroidal particle is $ 703 {\mu}_{\rm{B}}$. For disordered case, calculated value for spherical and spheroidal geometries are $68{\mu}_{\rm{B}}$ and $122{\mu}_{\rm{B}}$, respectively. Thus for $15\rm{nm}$ size particles, magnetic moment calculations using ordered surface and spheroidal geometry is very close to the experimental value $700 {\mu}_{\rm{B}}$.\cite{kodama} The experimental value of magnetic moment for $8.5\rm{nm}$ particle size \cite{tiwari1} at $10 K$ $(700 \mu_{\rm{B}})$ compares quite well with our calculated value $802 {\mu}_{\rm{B}}$ with ordered-surface and spherical geometry. However, magnetic moment value using particle with ordered-surface and spheroidal geometry is $550 \mu_{\rm{B}}$, a little less than the experimental value. In our model, the total magnetic moment per particle may slightly increase or decrease depending on a distribution of particle sizes and oblateness in the shape of particles. 

In Fig. \ref{magdist}(b) we have shown the total magnetic moment averaged over a lognormal distribution defined above as a function of the size of the particle. For an ordered case starting from very small sizes, we find that the magnetic moment increases with increasing the particle size and  attains a peak, and further increasing the particle size decreases the net magnetic moment. The averaged magnetic moment for a spheroid particle with $\epsilon=4$ has also been shown in the Fig. \ref{magdist}(b). For a spherical particle magnetic moment has peak value around $12 a_0$ whereas for spheroid case it is around $9a_0$.
As we can see from Fig. \ref{delrs}, the surface roughness effect is stronger only for particles of intermediate sizes, which confirms the greater role of surface roughness for these sizes. But for the nanoparticles of sizes greater than $12 a_0$, the surface roughness shell $\Delta$ becomes very small and a net magnetic moment arises largely due to the uncompensation of bulk N\'eel-state ordering which will tend to zero for large enough sizes showing the bulk character.
Thus Fig. \ref{magdist}(b) reflects a trend for ordered surface particles, where the net magnetic moment is very small for smallest size particles. Increasing the size, magnetic moment increases and reaches a maximum due to the greater role of surface, and again decreases towards the bulk value. 
In the same figure Fig \ref{magdist}(b), we have shown total magnetic moment for a disordered case averaged over the same distribution as discussed above. The magnetic moment in disordered case is showing an increasing trend with particle size, but the value is small as compared to ordered case. 
The magnetic moment values calculated here depends on the model parameters the core interaction parameter $\alpha$, and the surface roughness parameter $\beta$, and the width of the size distribution $\sigma$. Adjusting these parameters for a given size distribution and varying oblateness leads to the net magnetic moment value comparable to the experimental values. Tuning of the parameters $\alpha$ and $\beta$ are directly related to the multi-sublattice ordering for nanoparticles as predicted by Kodama {\it et al}.\cite{kodama}

\section{Conclusions}
\label{conclusion}
We have investigated finite-size and surface roughness effects in NiO nanoparticles. 
We have
found that the net magnetic moment due to finite-size fluctuations is nonmonotonic, oscillatory
and proportional to the particle size $R$, hence magnetization goes 
as $1/R^2$. The geometry of the particle also plays an important role in net magnetic moment. An oblate spheroid shape particle shows an increase in net magnetic moment by increasing oblateness of the particle. The experimental magnetic moments for various sizes are quite large compared to the
magnetic moments that arise as a finite-size fluctuation. The surface effects become very important in nanoparticles. We have incorporated surface effects in two different ways; an ordered surface where all the spins lying in the surface roughness shell are aligned due to internal field from the core spins, and a disordered surface where spins are randomly oriented in the surface roughness shell. Due to roughness of the surface and 
structural disorders, the uncompensated surface spins can be more easily deviated from the  
antiferromagnetic alignment by a magnetic field. 
We have introduced a surface anisotropy term and a ferromagnetic exchange interaction term for the surface spins, and a ferromagnetic exchange interaction term between core and surface spins lying at the interface of 
core and surface along with the bulk model in the Hamiltonian. A variational approach has been taken to find the dependence of the shell thickness on the size of particle. We have found that for nontrivial values of $\Delta$, the core interaction strength is modified which shows a signature of multi-sublattice ordering rather than two-sublattice ordering for smaller sizes.
$\Delta $ is showing size dependent fluctuations, with an envelope decay 
$\Delta\sim R^{-1/5}$.
We have shown that the total magnetic moment calculated with ordered as well as disordered surfaces displays size dependent fluctuations. For an ordered surface case, smoothening these fluctuations by a window-averaging, using a lognormal size distribution of nanoparticles, results a magnetic moment per particle which is very close to observed
experimental values of various sizes. We have also found that due to surface roughness effect, the net magnetic moment shows a trend where magnetic moment is very small for smallest size particles. Increasing the size, magnetic moment increases and reaches a maximum at $R\sim 12 a_0$ (depending upon the distribution of sizes and oblateness of the particles), and again decreases towards the bulk value.

\section*{Acknowledgments} 
We acknowledge K. P. Rajeev and S. D. Tiwari for extensive discussions. S.K.M  acknowledges 
D. D. B. Rao for helpful discussions, and the financial support provided by the  Council of 
Scientific and Industrial Research CSIR, Government of India.

\appendix{Spheroidal nanoparticles}
In this appendix, we calculate the total number of spins and net magnetic moment of a nanoparticle using spheroid geometry with polar radius $R/\epsilon$ and equatorial area $R$. 
We can write the three-dimensional position vector of the lattice sites in $l^{th}$ plane labelled by integers $m$, $n$ and $l$  is 
(using the 
cartesian unit vectors $ \hat{i}, \hat{j}, \hat{k} $) 
as
\begin{eqnarray}
 \vec{r}_{mnl}=(m+\frac{n}{2})a\hat{i}+\frac{\sqrt{3}n}{2}a\hat{j}+l \delta \hat{k}.
\end{eqnarray}
The spheroidal geometry implies that only those spins whose locations satisfy  ${x}_{mnl}^2+{y}_{mnl}^2+\epsilon^2{z}_{mnl}^2\leq R^2$, should be counted.

If $N_0$, $N_1$ and $N_2$ are the total number of lattice sites, counting from the planes of type A, B and C  respectively following section \ref{finite}, then the total number of lattice sites within a spheroid of equatorial radius $R$ will be  given as,
\begin{eqnarray}
 N_{\rm{spheroid}}(R)=\displaystyle\sum_{I=0}^{2} N_I(R),
\label{nelip}
\end{eqnarray}
where,
\begin{eqnarray}
N_I(R) = \displaystyle\sum_{\underset{ l\equiv I\left(Mod 3\right) }{mnl} }\Theta\left( R^2- ({x}_{mnl}-x_I)^2- ({y}_{mnl}-y_I)^2- \epsilon^2{z}_{mnl}^2 \right),
\label{eqnelip}
\end{eqnarray}
and $x_0=0$, $y_0=0$; $x_1=a/2$, $y_1=a/2\sqrt{3}$; and $x_2=a/2$, $y_2=-a/2\sqrt{3}$.
We transform the above equation using the Poission sum formula \cite{lighthill} as
\begin{eqnarray}
N_{I}(R) &=& \frac{1}{a^3} \sum_{p,q,w} \int e^{2\pi i (x' p + y' q + z' w)}\nonumber\\
 &\times& \Theta\left( R-\sqrt{({x'}_{mnl}-x_I)^2- ({y'}_{mnl}-y_I)^2- \epsilon^2{z'}_{mnl}^2 }\right) dx'dy'dz'.
\label{numelip}
\end{eqnarray}
Thus, the total number of spins can be written as
\begin{eqnarray}
N_{spheroid}(R)=\frac{16 \pi}{3\epsilon}\left( \frac{R}{a_0}\right)^3\displaystyle\sum_{I=0}^{2} \sum_{\lbrace\vec{G}\rbrace} \cos\left( \vec{G}.\vec{R}_I\right)  \frac{j_1\left( GR \right)}{G R},
\label{1nspelip}
\end{eqnarray}
 where $\vec{G}=\frac{4\pi}{\sqrt{3} a}\left[p \frac{\sqrt{3}}{2}\hat{i}+(q-\frac{p}{2}) \hat{j}+\frac{w}{2\sqrt{2}\epsilon}\hat{k}\right]$ is a three dimensional reciprocal lattice vector, labeled by three integers $p$, $q$ and $w$, and $\vec{R}_I=\vec{r}_I+I\epsilon \delta \hat{k}$.
 $j_1$ is a spherical Bessel function of order one \cite{abramowitz}. 

We can write the total magnetic moment for the ellipsoidal particle as
\begin{eqnarray}
M_{\rm{spheroid}}(R) = 2 {\mu}_{\rm{B}} \displaystyle\sum_{I=0}^{2} M_I(R),
\label{mnumelip}
\end{eqnarray}
where,
\begin{eqnarray}
M_I(R)= \displaystyle\sum_{\underset{l\equiv I\left(Mod 3\right) }{\vec{r}_{mnl}}} \left(-1 \right) 
^{l}N_{I}(R_l),
\label{mnumappend}
\end{eqnarray}
and $R_l=\sqrt{R^2-l^2\delta^2\epsilon^2}$.
 Applying the Poisson sum formula and proceeding analogously as we did above,
we calculate $M_I(R)$ and thus the total magnetic moment in the units of Bohr magneton $(\mu_{\rm{B}})$  as
\begin{eqnarray}
M_{\rm{spheroid}}(R) &=& \frac{32\pi}{3\epsilon} \frac{ R ^{2}}{{a_0}^3}\displaystyle\sum_{I=0}^{2}
\displaystyle\sum_{\lbrace\vec{g},w\rbrace}\cos\left( \vec{g}.{\vec{r}}_I\right) \nonumber\\
&\times&\cos \left( \frac{2 I \pi w}{3}\right)
\biggl( G^2 + \frac{3{\pi}^2}{{\epsilon}^2}
+2\frac{\sqrt{3}\pi G_z}{\epsilon}\biggr )^{-\frac{1}{2}} \nonumber\\
&\times& j_1\left\lbrace R\left( G^2+\frac{3{\pi}^2}{{\epsilon}^2} 
+\frac{2\sqrt{3}\pi 
G_z}{\epsilon}\right )^{\frac{1}{2}} \right\rbrace,
\label{magappend}
\end{eqnarray}
where
$\vec{G}=\vec{g}+G_{z} \hat{k}$ and $ G_z = \frac{2 \pi}{3\epsilon \delta} w$.


\begin{thebibliography}{45}
\bibitem{stamp} P.~C.~E.~Stamp, E.~M.~Chudnovsky, and B.~Barbara, {\it Int. J. Mod. Phys. B} {\bf 6}, 1355 (1992).
\bibitem{fiorani} D.~Fiorani, \textit{Surface Effects in Magnetic Nanoparticles} (Springer, New 
York, 2005).
  \bibitem{kodama} R.~H.~Kodama, Salah A.~ Makhlouf, and A.~E.~Berkowitz, {\it Phys. Rev.
  Lett.} {\bf 79}, 1393 (1997).
\bibitem{kodama1} R.~H.~Kodama, A.~E.~Berkowitz, {\it Phys. Rev. B} {\bf 59}, 6321 (1999). 

\bibitem{winkler} E.~Winkler, R.~D.~Zysler, M.~Vasquez Mansilla, and D.~Fiorani, {\it Phys. Rev. B} {\bf 72}, 132409 (2005).
\bibitem{tiwari} S.~D.~Tiwari and K.~P.~Rajeev, {\it Phys. Rev. B} {\bf 72}, 104433 (2005).
\bibitem{yi} J.~B.~Yi, J.~Ding, Y.~P.~Feng, G.~W.~Peng, G.~M.~Chow, Y.~Kawazoe, B.~H.~Liu, J.~H.~Yin, and S.~Thongmee, {\it Phys. Rev. B} {\bf 76}, 224402 (2007).
\bibitem{winkler1} E.~Winkler, R.~D.~Zysler, M.~Vasquez Mansilla, D.~Fiorani,  D.~Rinaldi, M.~Vasilakaki and K.~N Trohidou, {\it Nanotechnology} {\bf 19}, 185702 (2008).
\bibitem{morales} M.~A.~Morales, R.~Skomski, S.~Fritz, G.~Shelburne, J.~E.~Shield, Ming Yin, Stephen O'Brien, and D.~L.~Leslie-Pelecky, {\it Phys. Rev. B} {\bf 75}, 134423 (2007).
 \bibitem{behl}S.~M\o{}rup,D.~E.~Madsen, C.~Frandsen C.~R.~H.~Bahl and M.~F.~Hansen, {\it J.~Phys.: Condens. Matter} {\bf 19}, 213202 (2007).
\bibitem{tomou} A.~Tomou, D.~Gournis, I.~Panagiotopoulos, Y.~Huang, G.~C.~Hadjipanayis, and B.~J.~Kooi, {\it J. Appl. Phys.} {\bf 99}, 123915 (2006).
\bibitem{bhowmik} R.~N.~Bhowmik, R.~Nagarajan, and R.~Ranganathan {\it Phys. Rev. B} {\bf 69}, 054430 (2004). 
\bibitem{jagodic} M.~Jagodi\u{c}, Z.~Jagli\u{c}i\'{c}, A.~Jelen, Jin Bae Lee, Young-Min Kim, Hae Jin Kim and J Dolin\v{s}ek  {\it J. Phys.: Condens. Matter} {\bf 21}, 215302 (2009).

\bibitem{shull} C.~G.~Shull, W.~A.~Strauser, and E.~O.~Wollan, {\it Phys. Rev.} {\bf 83}, 333 (1951).

\bibitem{richard1} J.~T.~Richardson And W.~O.~Milligan, {\it Phys. Rev.} {\bf 102}, 1289 (1956).

\bibitem{makhlouf} S.~A.~Makhlouf, F.~T.~Parker, F.~E.~Spada and A.~E.~Berkowitz, {\it J. Appl. Phys.} {\bf 81}, 5561 (1997).
\bibitem{neel} L.~N\'eel in {\it Low Temperature Physics}, ed. C.~DeWitt, B.~Dreyfus and 
P.~G.~DeGennes, (Gordon and Beach, London, 1962), p.~411.
\bibitem{richard2}  J.~T.~Richardson, D.~I.~Yiagas, B.~Turk, K.~Forster, and M.~V.~Twigg, {\it J. Appl. 
Phys.} {\bf 70}, 6977 (1991).
\bibitem{schuele} W.~J.~Schuele And V.~D.~Deetscreek, {\it J. Appl. Phys.} {\bf 33}, 1136 (1962).
\bibitem{jacob} I.~S.~Jacobs and C.~P.~Bean, in {\it Magnetism}, ed. G.~T.~Rado and 
H.~Suhl, (Academic Press, New York, 1963) Vol. III, p.~294.
\bibitem{trohidou} X.~Zianni and K.~N.~Trohidou, {\it J. Appl. Phys.} {\bf 85}, 1050 (1999).
\bibitem{labarta} O.~Iglesias and A.~Labarta, {\it Phys. Rev. B} {\bf 63}, 184416 (2001).
 \bibitem{labarta1} O.~Iglesias and A.~Labarta, {\it Physica B} {\bf 343}, 286 (2004).
 \bibitem{labarta2} A.~Labarta, X.~Batlle and O.~Iglesias in {\it Surface Effects in
 Magnetic Nanoparticles}, ed. D.~Fiorani, (Springer, 2005) p.~105.

 \bibitem{labarta3} O.~Iglesias and A.~Labarta, {\it J. Magn. Magn. Mater.} {\bf 290}, 738 (2005).

\bibitem{rooksby} H.~P.~Rooksby, {\it Acta Cryst.} {\bf 1}, 226 (1948). 

 \bibitem{slack} G.~A.~Slack, {\it J. Appl. Phys.} {\bf 31}, 1571 (1960).

\bibitem{bartel} L.~C.~Bartel and B.~Morosin, {\it Phys. Rev. B}  {\bf 3}, 1039 (1971).

\bibitem{roth1}  W.~L.~Roth, {\it Phys. Rev.} {\bf 111}, 772 (1958).

\bibitem{roth2} W.~L.~Roth and G.~A.~Slack, {\it J. Appl. Phys.} {\bf 31}, 352S (1960).

\bibitem{roth3} W.~L.~Roth {\it J. Appl. Phys.} {\bf 31}, 2000 (1960).

\bibitem{hutchings}  M.~T.~Hutchings and E.~J.~Samuelsen, {\it Phys. Rev. B} {\bf 6}, 3447 (1972).

 \bibitem{behl1} C.~R.~H.~Bahl, M.~F.~Hansen, T.~Pedersen, S.~Saadi, K.~H.~Nielsen
B.~Lebech~and~S.~M\o{}rup {\it J.~Phys.: Condens. Matter} {\bf 18}, 4161 (2006).

\bibitem{mita} Y.~Mita, Y.~Ishida, M.~Kobayashi and S.~Endo  {\it J.~Phys.: Condens. Matter} {\bf 14}, 11173 (2002).
\bibitem{abramowitz} M.~Abramowitz and I.~A.~Stegun, {\it Handbook of Mathematical Functions}, (New york Dover 1972).
\bibitem{lighthill} M.~J.~Lighthill, {\it Fourier Analysis and Generalized Functions}, (Cambridge Univ. Press, London, 1958), p.~67-71.
\bibitem{zysler} R.~D.~Zysler, E.~Winkler, M.~Vasquez Mansilla and D.~Fiorani, {\it Physica B: Condensed
Matter} {\bf 384}, 277 (2006). 
\bibitem{buhrman} C.~G.~Granqvist and R.~A.~Buhrman {\it J. Appl. Phys.} {\bf 47}, 2200 (1976).
\bibitem{tiwari1} S.~D.~Tiwari and K.~P.~Rajeev {\it unpublished}.
\end{thebibliography}
\end{document}